\PassOptionsToPackage{prologue,dvipsnames}{xcolor}
\documentclass{article}

    \PassOptionsToPackage{numbers, compress, sort}{natbib}

\usepackage[preprint]{neurips_2024}

\usepackage[utf8]{inputenc} 
\usepackage[T1]{fontenc}    
\usepackage{hyperref}       
\usepackage{url}            
\usepackage{booktabs}       
\usepackage{amsfonts}       
\usepackage{nicefrac}      
\usepackage{microtype}      
\usepackage{xcolor}         
\usepackage{amsmath}
\usepackage{amssymb}
\usepackage{graphicx}
\usepackage{multirow}
\usepackage{pifont}
\usepackage[dvipsnames]{xcolor}
\usepackage{caption}
\usepackage{subcaption}

\title{KAD: No More FAD! An Effective and Efficient Evaluation Metric for Audio Generation}

\author{%
  Yoonjin Chung\thanks{Equal contribution} $^1$,
  Pilsun Eu$^{*1}$,
  Junwon Lee$^{2}$,
  Keunwoo Choi$^{1,3}$,\\
  \textbf{Juhan Nam$^{2}$,
  Ben Sangbae Chon$^{1}$}\\
  $^1$Gaudio Lab Inc., $^2$KAIST,
  $^3$Genentech
}

\begin{document}

\maketitle

\begin{abstract}
Although being widely adopted for evaluating generated audio signals, the Fréchet Audio Distance (FAD) suffers from significant limitations, including reliance on Gaussian assumptions, sensitivity to sample size, and high computational complexity. As an alternative, we introduce the Kernel Audio Distance (KAD), a novel, distribution-free, unbiased, and computationally efficient metric based on Maximum Mean Discrepancy (MMD). Through analysis and empirical validation, we demonstrate KAD’s advantages: (1) faster convergence with smaller sample sizes, enabling reliable evaluation with limited data; (2) lower computational cost, with scalable GPU acceleration; and (3) stronger alignment with human perceptual judgments. By leveraging advanced embeddings and characteristic kernels, KAD captures nuanced differences between real and generated audio. Open-sourced in the \texttt{kadtk}\footnote{\url{https://github.com/YoonjinXD/kadtk}} toolkit, KAD provides an efficient, reliable, and perceptually aligned benchmark for evaluating generative audio models.

\end{abstract}


\section{Introduction}\label{sec:intro}
\begin{table}[h]
\centering
\resizebox{0.65\textwidth}{!}{%
\begin{tabular}{@{}lccc@{}}
\toprule
    & Distribution-free?                                                 & Bias-free?  & Computation Cost\\ \midrule
FAD \cite{fad} & \textcolor{red}{\ding{56}} & \textcolor{red}{\ding{56}} & $\mathcal{O}(dN^2+d^3)$      \\
KAD (Ours) & \textcolor{ForestGreen}{\ding{52}} & \textcolor{ForestGreen}{\ding{52}} & $\mathcal{O}(dN^2)$      \\ \bottomrule
\end{tabular}%
}
\vspace{0.2cm}
\caption{Comparison of FAD and KAD: KAD is distribution-free, unbiased, and computationally efficient, even for high embedding dimensions ($d \leq 2048$) and large sample sizes ($N \leq 10k$).}
\label{tab:fad_kad_comparison}
\end{table}

As the demand for neural audio generation continues to grow across various domains such as content creation and virtual environments, innovative models are emerging to address a wide range of tasks. These include generating audio from textual descriptions, visual inputs, temporal data, or other audio signals, underscoring the importance of models that can process diverse types of inputs. Consequently, the need for robust and reliable methods to evaluate the quality of these models is becoming increasingly critical.

The Fréchet Audio Distance (FAD)~\cite{fad} is a widely used metric for evaluating the overall performance of audio generation models, measuring the dissimilarity between the statistical distributions of real and generated audio samples.
FAD is considered a simple yet effective measure for objective evaluation, making it a popular choice for assessing generative audio models across various tasks.

However, FAD has significant shortcomings that limit its effectiveness as a benchmark for generative audio models. 
First, it relies on the assumption that audio embeddings follow a Gaussian distribution, which often does not apply to real-world audio data with complex and diverse characteristics. Second, FAD suffers from an inherent bias in finite-sample estimation, which produces unreliable results particularly with smaller datasets. Finally, the computational cost of FAD is substantial, due to its high dependence on the embedding dimension size. 
These issues collectively challenge the practical use of FAD in evaluating modern generative audio models~\cite{jayasumana2024rethinking,chong2020effectively,tailleur2024correlation,gui2024adapting}.

\begin{figure}[t]
    \centering
    \includegraphics[width=\linewidth]{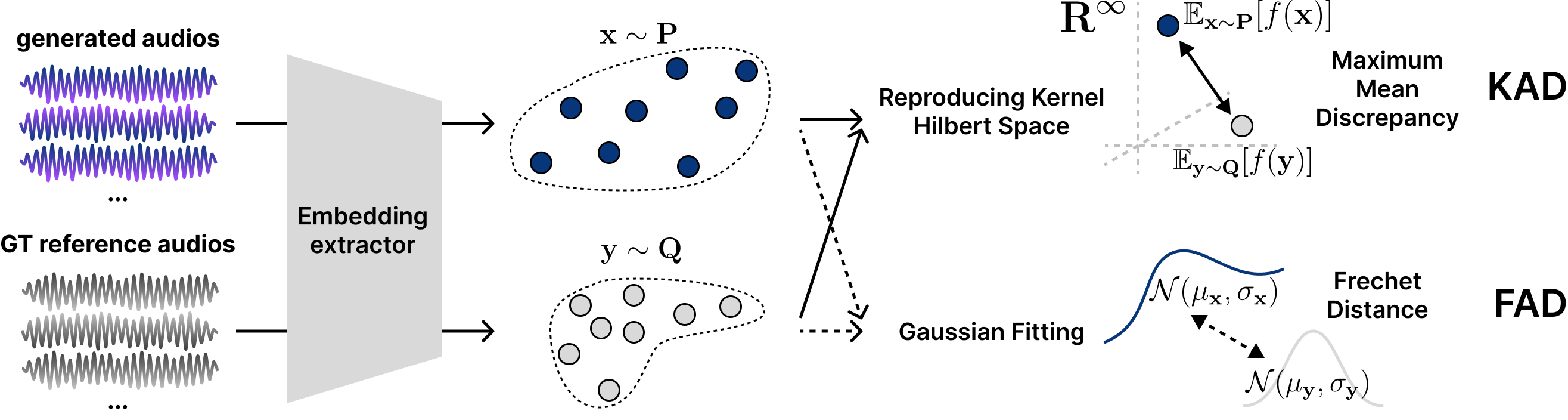}
    \caption{Comparison between KAD (Kernel Audio Distance) and FAD (Fréchet Audio Distance). KAD is a distribution-free metric that does not require any underlying assumptions for embedding distributions $P$ and $Q$.}
    \label{fig:kad&fad}
\end{figure}

Motivated by these limitations of FAD, we propose the Kernel Audio Distance (KAD), a novel evaluation metric for audio generation. The proposed metric leverages the Maximum Mean Discrepancy (MMD), a non-parametric measure that makes no assumptions about the underlying distribution of sample embeddings such as the normality assumption in FAD. A comparison between KAD and FAD is illustrated in Figure \ref{fig:kad&fad}.
In summary, our contributions are:
\begin{itemize}
    \item We propose KAD, a novel metric based on MMD for evaluating generative audio models.
    \item We provide empirical evidence demonstrating faster convergence with sample size, computational advantages, and stronger rank correlation with human evaluations of KAD.
    \item We provide guidelines to establish the practical applicability of KAD for selecting kernels, parameters, and embedding models to ensure consistent and reliable audio quality evaluation. 
\end{itemize}                      
The implementation of KAD is provided as an open-source toolkit named \path{kadtk} (for more detail, refer to Appendix \ref{sec:kadtk_release}).

\section{Related Works and Preliminaries}\label{sec:related}

\subsection{Fréchet Audio Distance and its Limitations}
The Fréchet Audio Distance (FAD)~\cite{fad} measures the difference between two sets of audio samples within their data embedding space. Specifically, it is an estimation of the Fréchet distance between the underlying distributions of two given embedding sample sets. FAD is an adaptation of the Fréchet Inception Distance (FID)\cite{fid} -- originally proposed for evaluating image generation models -- to the audio domain. The embeddings are typically extracted using an audio encoder model pretrained on real-world data such as VGGish\cite{vggish}, ensuring that the embeddings capture representative features of the audio samples for reliable evaluation. 

Given the ground-truth reference set embeddings $X = \{x_i\}_{i=1}^n$ and the target evaluation set $Y= ~ \{y_j\}_{j=1}^m$, FAD is defined by:
\begin{equation}
    \text{FAD}^2(X,Y) = \|\mu_X - \mu_Y\|_2^2 \;+\; \text{tr}\left(\Sigma_X + \Sigma_Y - 2\sqrt{\Sigma_X \Sigma_Y}\right),
\label{eq:FAD}
\end{equation}
where $X$ and $Y$ are assumed to be sampled from multivariate Gaussian distributions, fully characterized by their means $\mu_X, \mu_Y$ and covariances $\Sigma_X, \Sigma_Y$.

FAD is a conventional choice of metric for evaluating generative models in various domains, including text-to-audio~\cite{donahue2018adversarial,liu2023audioldm,huang2023makeanaudio,tango,audiogen,t-foley,audioldm2,consistencytta,tango2,tango_llm,auffusion,stableaudio_open,ezaudio} and vision-to-audio~\cite{comunita2024syncfusion,video-foley,rewas,frieren,maskvat,v-aura,vatt,multifoley,ssv2a,vintage,mmaudio,clipsonic,v2a-mapper,im2wav,foleygen,tiva} tasks, and is considered one of the standards for generative performance. Despite its popularity, FAD has three crucial limitations that undermine its efficacy and efficiency:
\begin{enumerate}
\item\textbf{Normality assumption}: The assumption that audio embeddings follow a Gaussian distribution often fails for real-world data. Such data are often asymmetrically distributed or unevenly clustered, which compromises the metric's ability to accurately measure similarity. Similar limitations have been observed in the computer vision domain, where the normality assumption was shown to be unsuitable for the Inception~\cite{inception-v3} embedding space used for FID~\cite{jayasumana2024rethinking}. Figure \ref{fig:clotho_umap} shows how the actual distribution of VGGish embeddings from the Clotho dataset (reduced via UMAP) differs significantly from the samples drawn from a Gaussian distribution with the same mean and covariance. This deviation highlights the limitations of the normality assumption for representing real-world audio data, leading to potential inaccuracies and over- or under-estimations in FAD values.

\begin{figure}[t]
    \centering
    \includegraphics[width=0.8\linewidth]{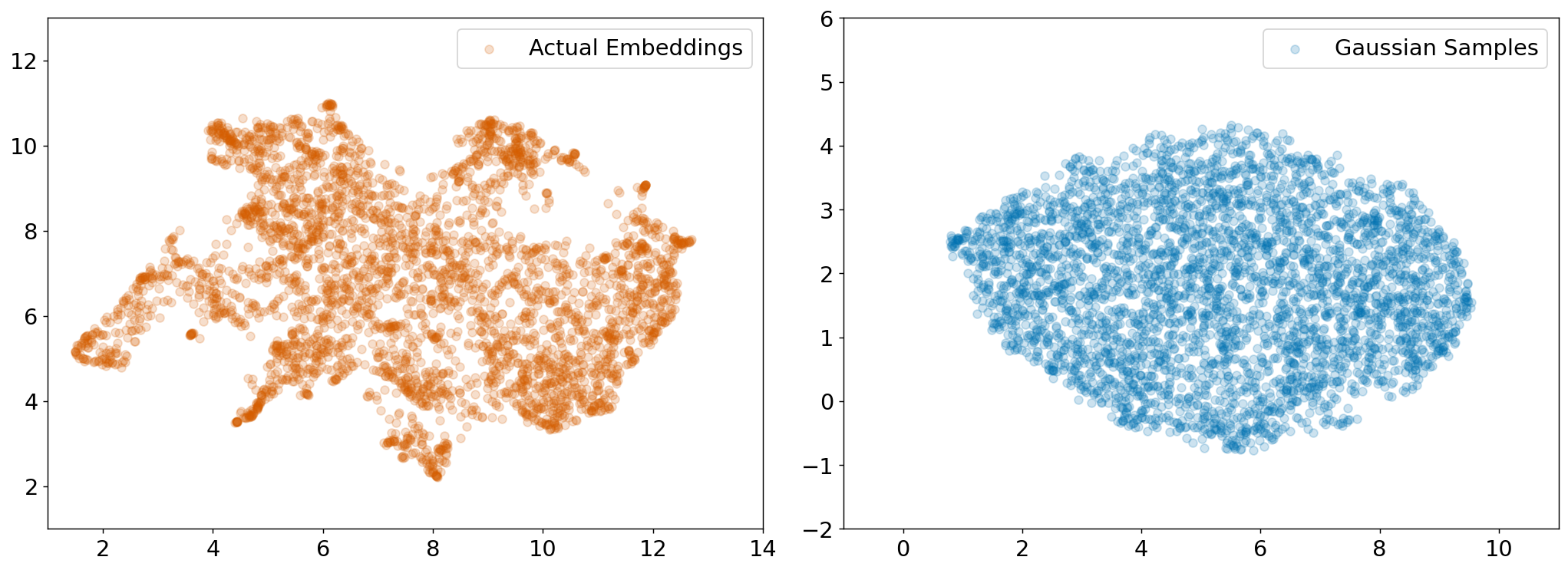}
    \caption{The left panel is the complex, non-Gaussian shape of the VGGish embeddings from the Clotho\cite{drossos2020clotho} train dataset, while the right panel depicts its assumed Gaussian distribution.}
    \label{fig:clotho_umap}
\end{figure}

\item\textbf{Sample size bias}: FAD is an inherently biased metric~\cite{gui2024adapting} which requires a large number of audio samples for a reliable result. Given the embedding sample size $N =\max(n,m)<\infty$, 
The bias in FAD from the finite-sample estimation of sample covariance increases as $\mathcal{O}(1/N)$ (for a detailed derivation, refer to Appendix~\ref{sec:appendix_fad_bias}). Similar findings have also been reported for FID~\cite{chong2020effectively}.
This necessitates the use of larger datasets for more accurate evaluations, which is particularly undesirable in the audio domain where high-quality data is relatively scarce compared to image datasets. Furthermore, this sample size bias creates the potential for manipulation: increasing $N$ can artificially reduce bias, leading to better FAD scores and the appearance of improved performance. Naturally, this further undermines the credibility of FAD as a reliable evaluation metric.

\item\textbf{High computational cost}: 
The time complexity of FAD is given by $\mathcal{O}(dN^2+d^3)$, which scales poorly with the number of dimensions of the embedding space. This makes the calculations cumbersome when using audio embedding models that produce high-dimensional embeddings. Moreover, the calculation of square-roots of covariance matrices in Equation \ref{eq:FAD} is not easily parallelized, limiting the utilization of parallel computing.

\end{enumerate}

\subsection{Maximum Mean Discrepancy} 
\label{sec:mmd}

To address the limitations of FAD, we adopt the Maximum Mean Discrepancy (MMD)~\cite{grettonmmd}. Originally proposed for a statistical test to distinguish whether two samples come from the same distribution, MMD is capable of capturing differences not only in mean and variance but also in higher-order moments. It is also distribution-free, meaning that it does not assume that the samples belong to a specific family of distributions (e.g. Gaussian). This allows for a more comprehensive comparison of how two sets of audio samples differ in their embedding spaces.

The MMD between two distributions $P$ and $Q$ is defined as: 
\begin{equation}
    \text{MMD}(\mathcal{F},P,Q) = \sup_{f \in \mathcal{F}} 
    \Bigl( \mathbb{E}_{x \sim P}[f(x)] \;-\; \mathbb{E}_{y \sim Q}[f(y)] \Bigr),
    \label{eq:mmd_def}
\end{equation}
where $\mathcal{F}$ is a class of functions chosen to detect differences between $P$ and $Q$. 

When $\mathcal{F}$ is chosen to be the Reproducing Kernel Hilbert Space (RKHS) induced by a kernel function $k(\cdot,\cdot)$, calculating the MMD corresponds to measuring the Euclidean distance between the mean embedding positions after mapping the data into a high-dimensional feature space. The high-dimensional mapping is necessary because it reveals the nonlinear differences between two distributions that may not be apparent in a lower-dimensional setting. 

Rather than computing these high-dimensional representations explicitly, kernel operations can be used to calculate the distances in the RKHS directly in the original embedding space. This technique, often referred to as the "kernel trick," allows the metric to leverage high-dimensional -- or even infinite-dimensional -- representations that would otherwise be infeasible to compute. With this setup, the MMD can be computed entirely through pairwise comparisons of the samples: 
\begin{equation}
    \text{MMD}^2(P, Q) = \mathbf{E}_{x,x'}[k(x,x')] + \mathbf{E}_{y,y'}[k(y,y')] - 2\,\mathbf{E}_{x,y}[k(x,y)],
    \label{eq:mmd_kernel}
\end{equation}
where $x, x'$ are drawn from $P$ and $y, y'$ are drawn from $Q$. 

For the finite samples in the reference set $X = \{x_i\}_{i=1}^n$ and the evaluation set $Y = \{y_j\}_{j=1}^m$, an unbiased estimator of \ref{eq:mmd_kernel} is: 
\begin{align}
\label{eq:MMD_unbiased}
    \widehat{\text{MMD}}^2_{\text{unbiased}}(X,Y) &= \frac{1}{n(n-1)}\sum_{i\neq j}k(x_i,x_j) \;+\; \frac{1}{m(m-1)}\sum_{i\neq j}k(y_i,y_j) \nonumber\\
    &\quad - \frac{2}{nm}\sum_{i=1}^n\sum_{j=1}^m k(x_i,y_j).
\end{align}

One of the most widely used MMD-based metrics for evaluating generative models is the Kernel Inception Distance (KID) proposed by Bińkowski et al.~\cite{binkowski2018demystifyingmmd}, as the squared MMD value between Inception embeddings of images with a cubic polynomial kernel. KID was used as an evaluation metric in several audio-related works~\cite{nistal2021comparing, nistal2024diff, grachten2024measuring, shi2024versa}, particularly in music generation.
For image quality, Jayasumana et al.~\cite{jayasumana2024rethinking} proposed the CLIP Maximum Mean Discrepancy (CMMD) for evaluation on CLIP embeddings~\cite{clip} and a Gaussian kernel, with Novack et al.~\cite{novack2024presto} following similar practices but on CLAP~\cite{clap_laion} embeddings.
Many of these works acknowledge the potential advantages of MMD. However, to our knowledge, there has been no comprehensive study of its reliability in comparison to FAD or the effect of various design choices.

\section{Kernel Audio Distance}\label{sec:kad}

In this section, we propose the \emph{Kernel Audio Distance} (KAD), a reliable and computationally efficient metric for evaluating audio generation models. 

The fundamental requirement for such a metric is its ability to capture perceptually meaningful differences between generated and reference audio. Provided that the embedding space sufficiently encodes these perceptually relevant features, a reliable metric must be capable of accurately comparing embedding distributions without imposing restrictive assumptions. To this end, we adopt the MMD, whose distribution-free nature eliminates the need for parametric assumptions and enables a comprehensive comparison between two embedding distributions.

We define KAD as follows:
\begin{equation}
    \text{KAD} = \alpha \cdot \widehat{\text{MMD}}^2_{\text{unbiased}},
\end{equation}
where $\alpha$ is a resolution scaling factor introduced for convenient score comparison. We set $\alpha = 100$ as the default.

\subsection{The Strengths of KAD}
Along with its robust theoretical foundation, KAD also provides key practical advantages over FAD:

\textbf{Unbiased Nature}: The KAD score is independent of the sample size, making it robust to smaller samples without employing bias-correction procedures. By contrast, the bias-correction for FAD (e.g., $\text{FAD}_\infty$~\cite{gui2024adapting}) relies on linear fitting of results at multiple sample sizes. This independence makes KAD especially robust in data-scarce conditions, such as early-stage evaluations of generative models or when high-quality reference datasets are limited.

\textbf{Overall Computational Efficiency}: KAD has a time complexity of $\mathcal{O}(dN^2)$. In practice, this can be significantly faster than $\mathcal{O}(dN^2 + d^3)$ for FAD, as the $d^3$ term can dominate with higher dimensionality. Therefore, KAD is more scalable for higher-dimensional embeddings typical of modern deep audio models.

\textbf{Parallel Computation} The pairwise operations in the computation of KAD (Eq.~\ref{eq:MMD_unbiased}), $- \frac{2}{nm}\sum_{i=1}^n\sum_{j=1}^m k(x_i,y_j)$, can be performed in parallel, enabling substantial acceleration. 


\subsection{Kernel Function and Bandwidth Selection}\label{subsec:kad_kernel}

KAD relies on evaluating pairwise relationships between embeddings through a kernel function. Many commonly used kernels such as Gaussian, Laplacian, and Matérn kernels are examples of a \textit{characteristic} kernel, meaning that the MMD it induces \textit{i)} fully distinguishes between two embedding distributions and \textit{ii)} is zero if and only if the two distributions under test are identical~\cite{sriperumbudur2011universality}. This property is crucial for model evaluation as it ensures that the KAD metric captures meaningful differences in the embedding distributions of real and generated audio samples. As an example, a cubic polynomial kernel $(x^Ty + 1)^2$ cannot differentiate between distributions with the same mean, variance and skewness, but different kurtosis~\cite{sriperumbudur2010hilbert}.

For the KAD, we choose the Gaussian radial basis function (RBF) kernel:
\begin{equation}
    k(\mathbf{x}, \mathbf{y}) = \exp\left(-\frac{\|\mathbf{x}-\mathbf{y}\|^2}{2\sigma^2}\right),
\end{equation}
where $\sigma$ is the bandwidth parameter. An implicit mapping $\phi(x)$ to the RKHS of the Gaussian RBF kernel is infinite-dimensional and is defined by the property $\langle \varphi(\mathbf{x}), \varphi(\mathbf{y}) \rangle = k(\mathbf{x}, \mathbf{y})$. This kernel has been extensively analyzed and validated in MMD applications~\cite{jayasumana2024rethinking, rustamov2019closed} for its smoothness, balanced sensitivity to both local and global variations, and well-studied performance across diverse datasets and modalities.

For the value of the bandwidth parameter $\sigma$, we follow the commonly adopted median distance heuristic~\cite{grettonmmd}, setting $\sigma$ as the median pairwise distance between the embeddings within the reference set. This heuristic provides a stable baseline with minimal tuning, ensuring the kernel is neither too flat (insensitive to dissimilarities) nor too peaked (over-sensitive to noise). While exploring adaptive or data-driven kernel selection is beyond the scope of this work, our initial experiments indicate that the median heuristic is sufficiently effective. More details are discussed in Appendix \ref{sec:appendix_bandwidth}.

\section{Experiments}\label{sec:exp_results}
In this section, we present our empirical findings on KAD and comparison with FAD across three key perspectives: (1) Alignment with Human Perception, (2) Convergence with Sample Size and (3) Computation Cost.

\begin{figure}[t]
    \centering
    \includegraphics[width=0.9\linewidth]{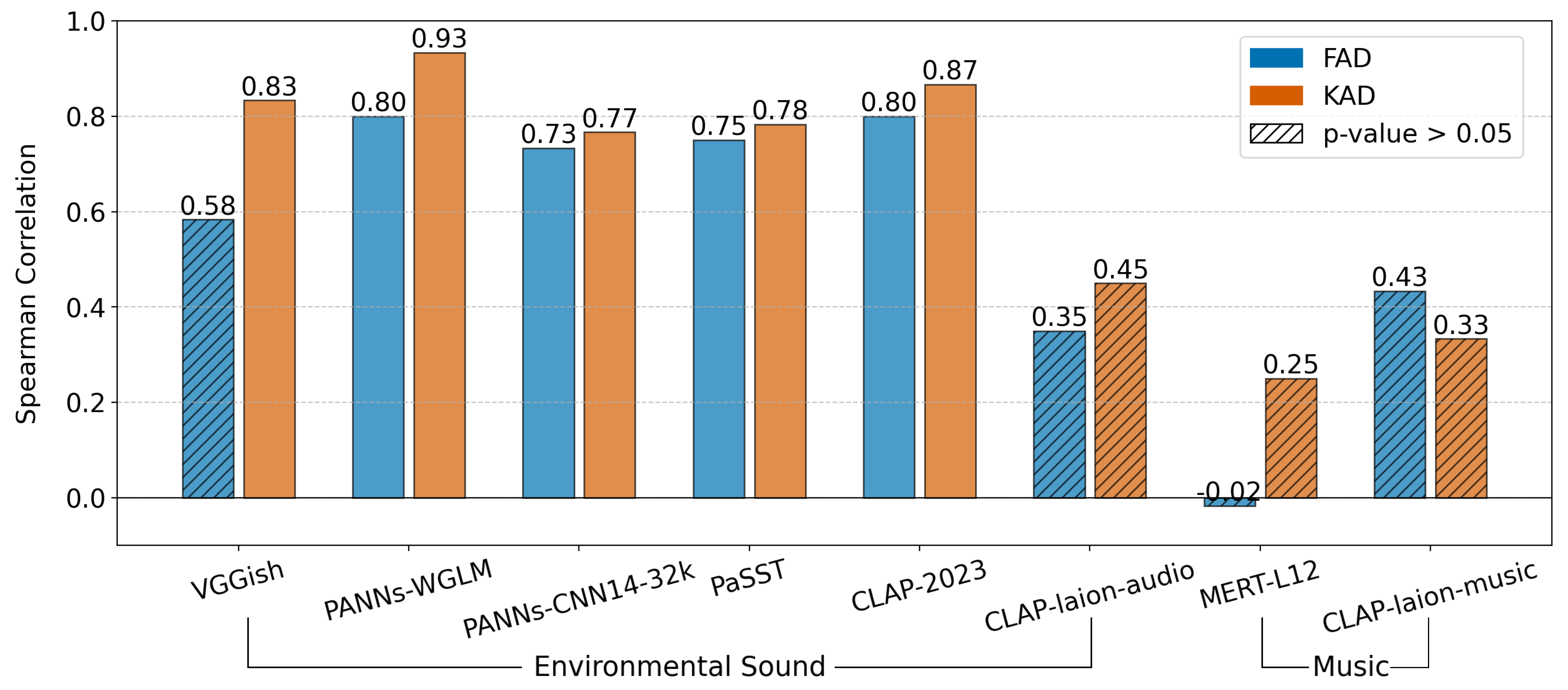}
    \caption{Spearman correlations between metric scores and human perceptual ratings for different embedding models. Since lower scores imply better results for both metrics, correlation values are negative. Correlation values are multiplied by -1 for the convenience of visualization. KAD (orange) consistently achieves higher alignment than FAD (blue).}
    \label{fig:human_alignment}
\end{figure}

\subsection{Experiment 1: Reliability of KAD in Perceptual Alignment}
\label{ssec:human_perception}
A reliable evaluation metric for generative audio should be closely aligned with the human perception of audio quality. While FAD relies on a normality assumption that may not accurately capture the multimodal nature of real-world audio embeddings, KAD takes a distribution-free approach. This flexibility allows it to handle complex acoustic feature representations and potentially align more closely with how humans perceive audio quality.

To validate the perceptual alignment of KAD in comparison with FAD, we use data from the DCASE 2023 Challenge Task 7 submissions~\cite{dcase2023,BaiJLESS2023,ChangHYU2023,ChonGLI2023,ChunChosun2023,ChungKAIST2023,GuanHEU2023,JungKT2023,KamathNUS2023,LeeMARG2023,Leemaum2023,PillayCMU2023,WendnerJKU2023,XieSJTU2023,YiSURREY2023,QianbinBIT2023,QianXuBIT2023,ScheiblerLINE2023} for Foley sound generation. This dataset provides human rating scores on audio quality for 9 different audio generation models, making it a reliable benchmark for correlating objective metrics with subjective judgments.

We compute both KAD and FAD using embedding from several well-known models, including VGGish~\cite{vggish}, PANNs~\cite{panns}, CLAP~\cite{clap_ms, clap_laion}, and PaSST~\cite{passt}, all of which are trained on environmental sounds. These embedding models are widely used for the calculation of FAD scores for text-to-audio \cite{donahue2018adversarial,liu2023audioldm,huang2023makeanaudio,tango,audiogen,t-foley,audioldm2,consistencytta,tango2,tango_llm,auffusion,stableaudio_open,ezaudio} and vision-to-audio generation\cite{comunita2024syncfusion,video-foley,rewas,frieren,maskvat,v-aura,vatt,multifoley,ssv2a,vintage,mmaudio,clipsonic,v2a-mapper,im2wav,foleygen,tiva}. Since music-focused models can differ substantially in their learned representations, we also include MERT~\cite{mert} and CLAP-laion-music~\cite{clap_laion} for completeness. We then measure the Spearman rank correlation between each metric's scores and the average human evaluation scores, as well as the p-value. Correlations with $p > 0.05$ are shaded in \ref{fig:human_alignment} to indicate a lack of statistical significance.

As shown in Figure \ref{fig:human_alignment}, KAD exhibits a Spearman correlation of up to $-0.93$, notably outperforming FAD whose strongest correlation is $-0.80$. This suggests that KAD is more effective for differentiating the perceptual nuances captured within the audio data embeddings from a wide range of common audio representations. In contrast, embeddings trained on music data (MERT and CLAP-laion-music) show weaker alignment, consistent with previous findings\cite{tailleur2024correlation}.

Among the tested embedding models, PANNs-WGLM(WaveGram-LogMel) achieves the strongest correlation with human judgments, aligning with prior research that highlighted its suitability for FAD-based evaluations~\cite{tailleur2024correlation}. Based on this observation, we select PANNs-WGLM as the primary embedding model in subsequent experiments to further investigate the performance of KAD.

\subsection{Experiment 2: Convergence with Sample Size}

To compare how KAD and FAD converge as the evaluation set size $N$ increases, we use the \textit{eval} split of the Clotho 2 dataset~\cite{drossos2020clotho} with 1045 samples as the reference set, and samples generated using AudioLDM~\cite{liu2023audioldm} as the evaluation set. The evaluation samples were generated by conditioning on text captions from the \textit{dev} split of the Clotho 2 dataset. The number of generated samples starts at $N=100$ and gradually increases up to $N=3839$ (the total size of the Clotho 2 \textit{dev} split). We compute both KAD and FAD under these varying $N$ values to observe their biases and convergence rates. 

\begin{figure}[t]
    \centering
    \includegraphics[width=0.45\textwidth]{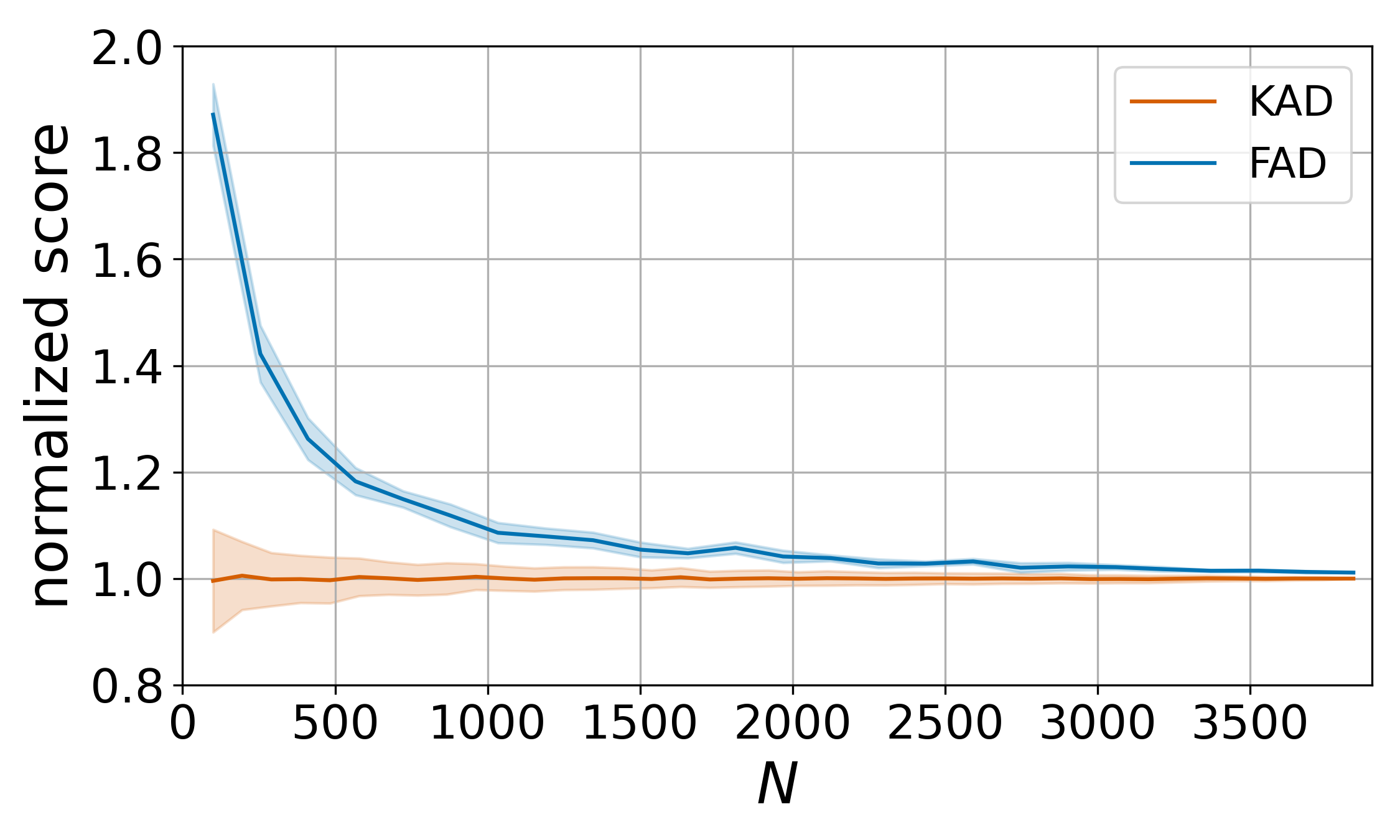}
    \vspace{-0.3cm}
    \caption{Normalized FAD and KAD scores against increasing embedding sample size. Scores are normalized by their respective extrapolated values at $N=\infty$. The shaded regions indicate standard deviations.}
    \label{fig:sample_convergence}
\end{figure}

Figure \ref{fig:sample_convergence} displays how KAD and FAD evolve as $N$ increases, normalizing each metric by its extrapolated value at $N=\infty$. At small $N$, FAD shows a distinct positive bias, deviating substantially from its stable value. This deviation decreases roughly by half whenever the sample size doubles, indicating that a large $N$ is needed for FAD to become reliable.

By contrast, KAD remains close to its asymptotic value even at relatively small $N$, reflecting its unbiased nature. While KAD does exhibit a relatively larger standard deviation (the shaded region) for smaller $N$, this uncertainty band narrows quickly. Notably, even when accounting for the standard deviation, the range of error for KAD is bounded by the magnitude of bias for FAD, up to the largest sample size tested ($N=3839$). These results show that KAD can serve as a more stable evaluation metric, especially when the availability of generated audio samples is limited.

\subsection{Experiment 3: Computation Cost Comparison}
\label{ssec:exp_compute}
To assess the computational efficiency of KAD relative to FAD, we measure their wall-clock times on both CPU and GPU across varying embedding dimensions $d$ and sample sizes $N$. We use PANNs-WGLM~\cite{panns}, VGGish~\cite{vggish}, and CLAP~\cite{clap_ms} -- encompassing dimension sizes from $d=128$ (VGGish) to $d=2048$ (PANNs-WGLM). The sample sizes range up to 10k to cover typical open-source audio-text datasets like Clotho~\cite{drossos2020clotho} and AudioCaps~\cite{kim2019audiocaps}.

For the measurements, AMD EPYC 7413 CPU (24 cores) and an Nvidia RTX 3090 GPU were used, and the code is implemented on PyTorch for both KAD and FAD calculations. For FAD, we refactored the Microsoft FAD toolkit~\cite{gui2024adapting} for consistency in CPU/GPU usage, thereby ensuring the comparability of runtime measurements. All values were calculated in single-precision floating points.

\begin{figure}[t]
    \centering
    \begin{subfigure}{0.48\textwidth}
        \centering
        \includegraphics[width=\linewidth]{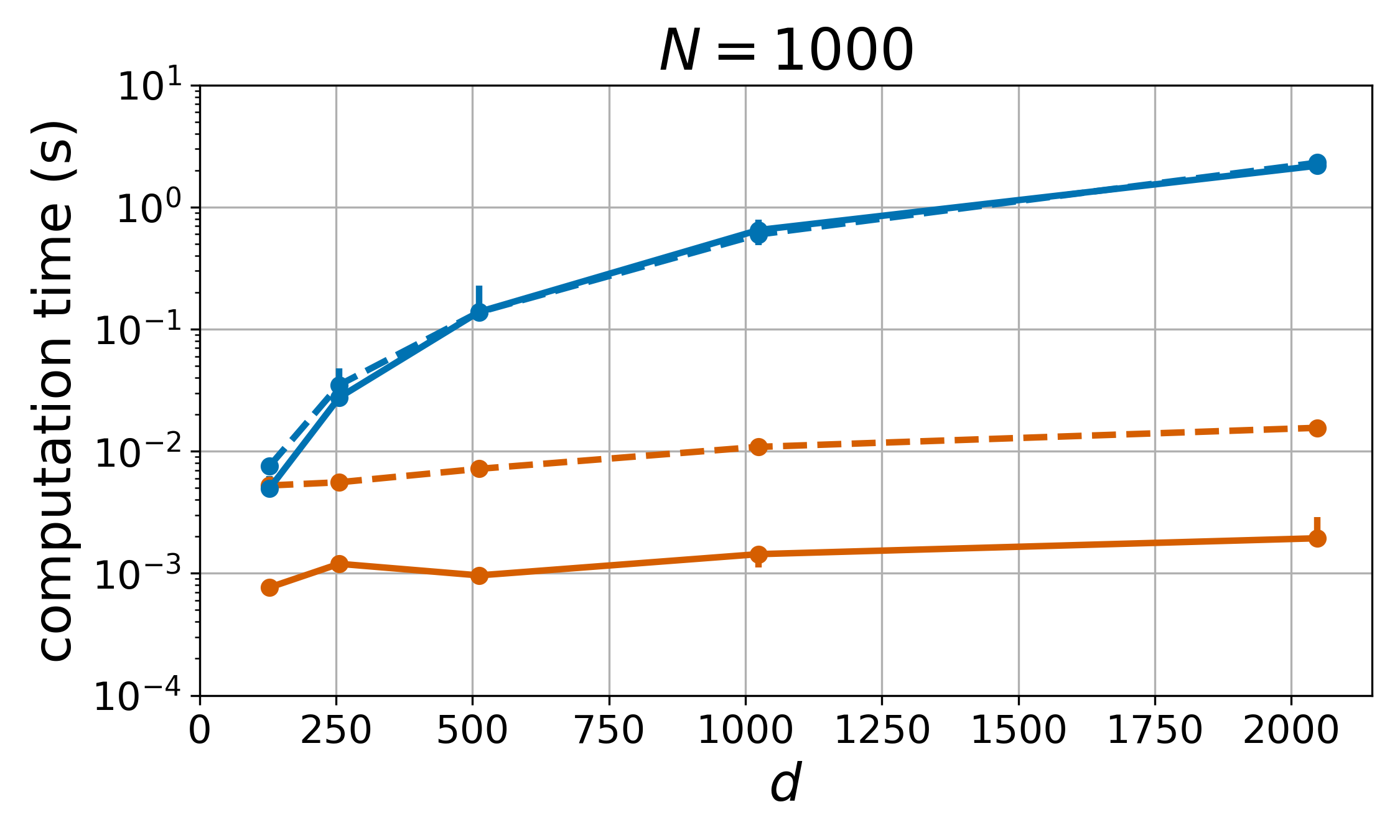}
        \vspace{-0.7cm}
        \subcaption{}
        \label{fig:computation_n1000}
    \end{subfigure}
    \begin{subfigure}{0.48\textwidth}
        \centering
        \includegraphics[width=\linewidth]{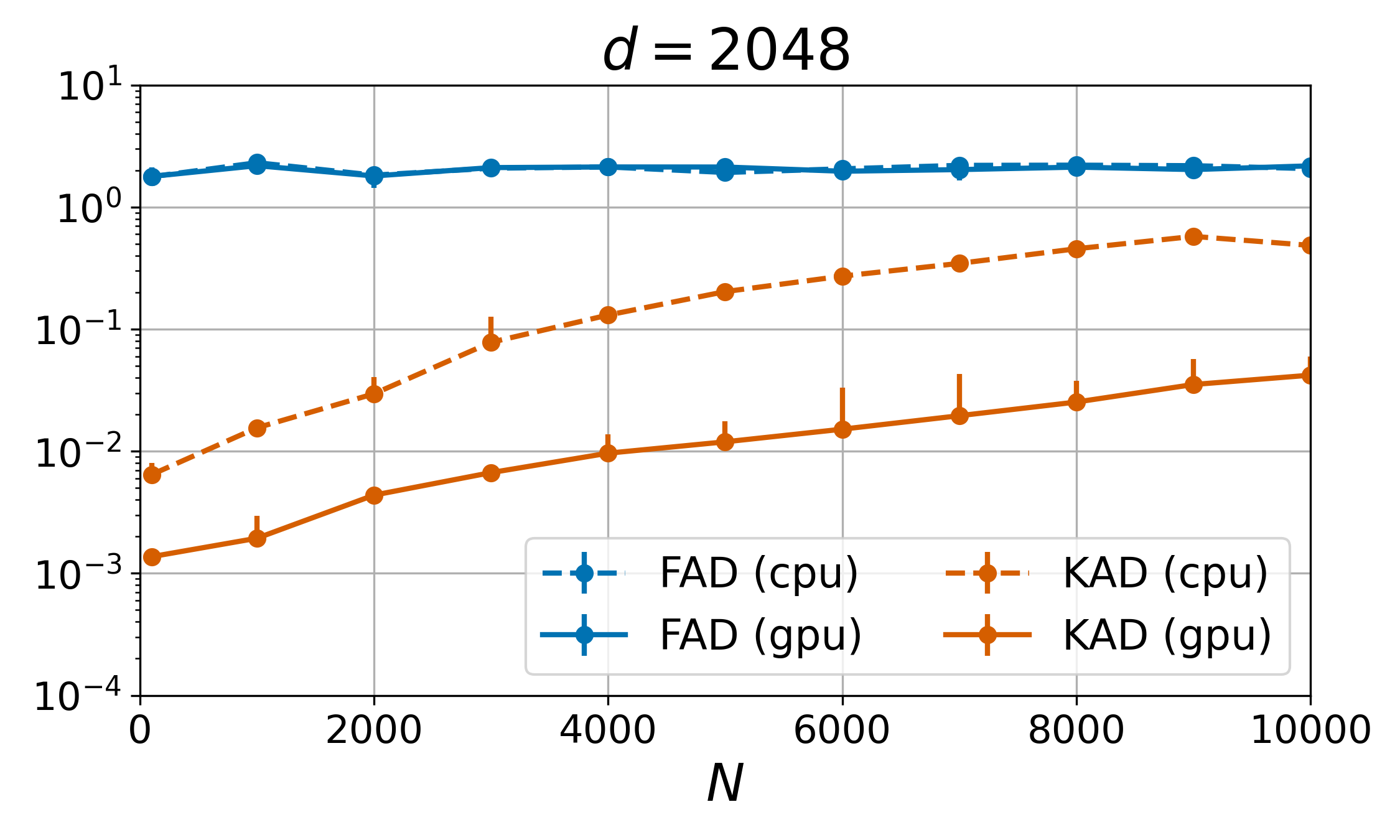}
        \vspace{-0.7cm}
        \subcaption{}
        \label{fig:computation_d2048}
    \end{subfigure}
    \vspace{-0.2cm}
    \caption{Comparison of FAD and KAD wall-clock computation times. 
    (a) $N=1000$ with varying $d$. (b) $d=2048$ with varying $N$. Solid lines indicate CPU usage and dotted lines indicate GPU usage. Error bars mark the 5th to 95th percentile of 200 trials.}
    \label{fig:computation}
\end{figure}

Figure~\ref{fig:computation} shows that FAD’s computation time increases dramatically with dimension size $d$, whereas KAD remains relatively stable. This stark difference aligns with the theoretical $d^3$ scaling of FAD, in contrast to KAD’s weaker dependence on $d$. FAD exhibits significant computational overhead at high dimensions even for small sample sizes. Figure~\ref{fig:computation_n1000} highlights how FAD’s wall-clock time (blue lines) escalates with $d$, while KAD (orange lines) remains nearly flat. At $d=2048$, the runtime gap can reach three orders of magnitude. Figure \ref{fig:computation_d2048} further confirms that the main bottleneck for FAD is dimension size, rather than the number of samples. This behavior indicates that FAD is less practical when evaluating embeddings with large $d$ or on resource-limited systems.

Furthermore, KAD benefits considerably from GPU acceleration (dotted vs. solid orange lines), achieving more than an order of magnitude of speedup. Table \ref{table:compute_comparison} quantifies these observations, showing consistent performance advantages of KAD over FAD under both CPU and GPU conditions.

\begin{table}[t!]
\centering
\resizebox{0.75\textwidth}{!}{%
\begin{tabular}{@{}ccrrrr@{}}
\toprule
\multirow{2}{*}{$d$}    & \multirow{2}{*}{$N$} & \multicolumn{2}{c}{CPU}                                                                               & \multicolumn{2}{c}{GPU}                                                                               \\
                        &                      & \multicolumn{1}{c}{KAD (ours)} & \multicolumn{1}{c}{FAD} & \multicolumn{1}{c}{KAD (ours)} & \multicolumn{1}{c}{FAD} \\ \midrule
\multirow{3}{*}{$128$}  & $100$                & 2.8 ± 0.06                                        & 5.7 ± 0.03                               & 0.6 ± 0.03                               & 5.4 ± 0.02                                        \\
                        & $5000$               & 102.8 ± 1.17                                        & 6.7 ± 0.09                                        & 4.1 ± 0.06                                        & 7.3 ± 0.12                                        \\
                        & $10000$              & 424.2 ± 4.00                                        & 6.9 ± 0.19                                        & 12.8 ± 0.10                                       & 7.9 ± 0.08                                        \\ \midrule
\multirow{3}{*}{$512$}  & $100$                & 2.8 ± 0.07                                        & 130.2 ± 0.65                               & 1.3 ± 0.01                               & 107.7 ± 0.29                                        \\
                        & $5000$               & 132.0 ± 1.37                                        & 155.5 ± 2.72                                        & 5.4 ± 0.12                                        & 128.5 ± 1.70                                        \\
                        & $10000$              & 461.5 ± 3.16                                        & 154.6 ± 2.65                                        & 17.3 ± 0.20                                       & 134.2 ± 1.83                                        \\ \midrule
\multirow{3}{*}{$2048$} & $100$                & 6.8 ± 0.12                                        & 1776.2 ± 14.5                              & 1.4 ± 0.03                               & 1829.1 ± 21.3                                       \\
                        & $5000$               & 204.6 ± 1.98                                        & 1921.1 ± 14.1                                       & 13.0 ± 0.41                                       & 2136.5 ± 21.8                                       \\
                        & $10000$              & 497.9 ± 4.35                                     & 2074.9 ± 20.5                                       & 46.3 ± 2.41                                       & 2174.4 ± 21.6                                       \\ \bottomrule
\end{tabular}%
}
\vspace{5pt}
\caption{Mean wall-clock times with 95\% confidence intervals over 200 trials, in milliseconds. KAD on GPU consistently runs faster than FAD for $N=100$ and $5000$, as well as for $N=10000$ at higher dimensions.}
\label{table:compute_comparison}
\end{table}

\section{Conclusion}\label{sec:conclusion}

In this paper, we addressed key limitations of the Fréchet Audio Distance (FAD) for evaluating generative audio models and proposed the Kernel Audio Distance (KAD) as a more robust alternative. Built on the Maximum Mean Discrepancy (MMD), KAD avoids making statistical assumptions about the embedding distributions, provides unbiased results for all sample sizes, and offers a computational complexity  that scale more efficiently, particularly at higher dimensionalities.

We define KAD as the MMD between reference and evaluation audio embedding sets using a Gaussian RBF kernel with the median-distance bandwidth heuristic. To validate its effectiveness, we compare both KAD and FAD against human evaluation data, observe their convergence behaviors with increasing sample sizes, and measure their CPU and GPU runtimes across a range of dimensionalities and sample sizes.

Our findings show that KAD aligns more strongly with human judgments than FAD across various common audio embedding models, with especially high correlation with PANNs-WGLM. Moreover, its score remains consistent regardless of sample size, making it practical for resource-constrained or early-stage model evaluations, and its computational overhead is up to orders of magnitude lower for higher dimensional (\textasciitilde2024) embeddings compared to FAD due to the reduction of the complexity from $\mathcal{O}(dN^2 + d^3)$ to $\mathcal{O}(dN^2)$ and its amenability to parallel computation.

These advantages position KAD as an efficient, comprehensive, and scalable tool for benchmarking generative audio models.By more accurately capturing human-perceived audio quality, KAD can support the development of more reliable evaluation practices in the field. The accompanying open-source toolkit is provided to encourage widespread adoption, experimentation, and ongoing improvements to the development and assessment of generative audio models.



{
\small
\bibliography{ref}
\bibliographystyle{unsrt}
}

\newpage

\appendix

\section{Median Pair-wise Distance Heuristic for the Kernel Bandwidth}
\label{sec:appendix_bandwidth}

\begin{figure}[ht]
    \centering
    \includegraphics[width=0.85\textwidth]{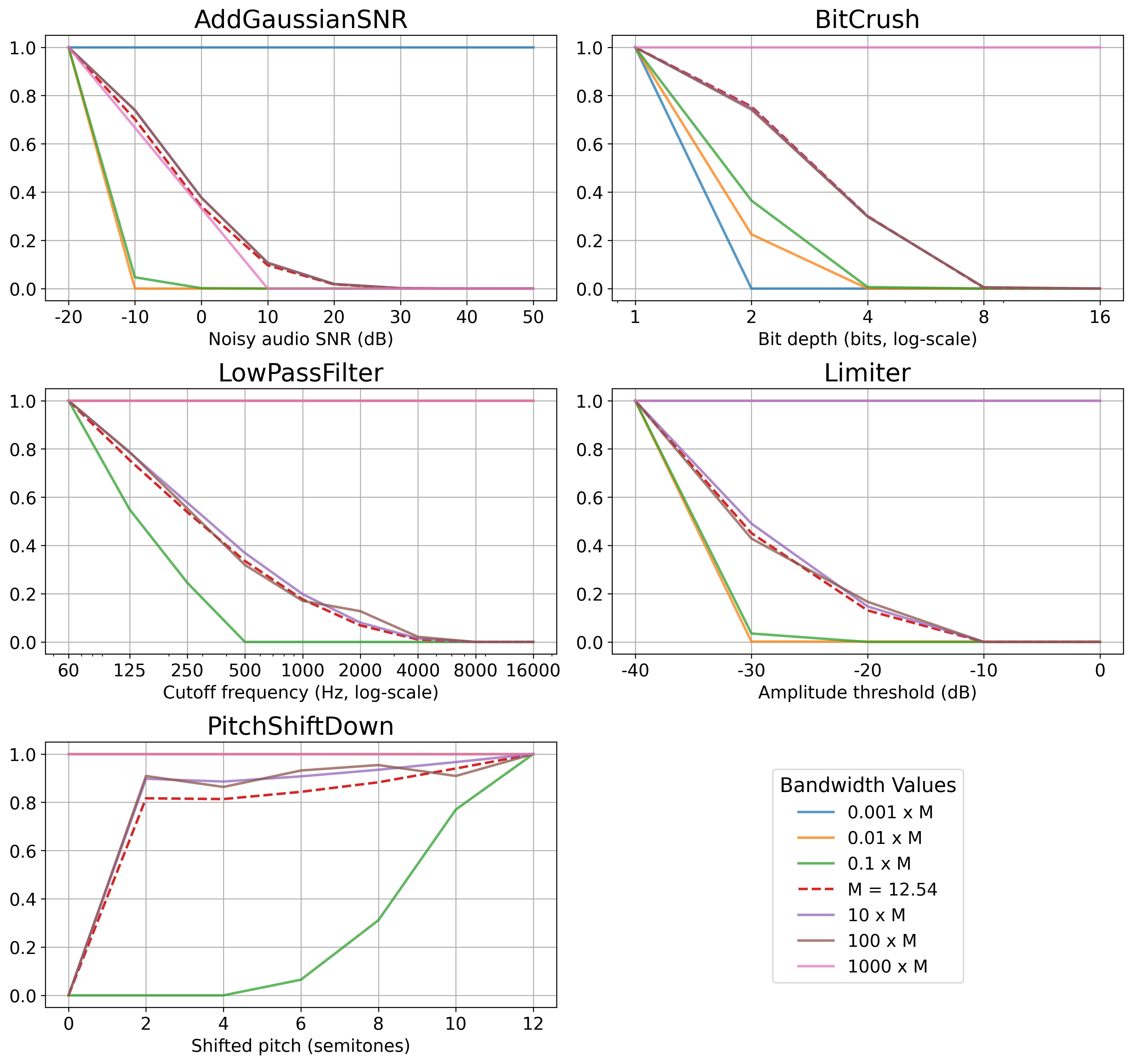}
    \caption{Effect of audio degradations to the MMD values, where the maximum value is normalized to 1 for each bandwidth result. MMD values that are smaller than $\varepsilon=10^{-12}$ are clipped to $\varepsilon$ before normalization.}
    \label{fig:bandwidth_score_norm}
\end{figure}

To assess the effectiveness of the median pairwise distance heuristic for setting the bandwidth parameter $\sigma$, we compared the MMD values calculated at different bandwidth settings as the quality of the evaluation data is artificially degraded. If the heuristic to works as intended, the MMD at the median distance bandwidth should be well-correlated with the audio quality for the perceptually relevant ranges of degradation.

We compared the MMD scores between the clean audio files from the Clotho dataset's \textit{eval} split, and the same files under various audio signal transformations including Gaussian noise injection, bit depth reduction, low-pass filtering, dynamic range reduction (limiter), and pitch shifts. The degradations were applied using the \textit{audiomentations} python library~\cite{audiomentations}.

For each transformation, MMD scores were calculated using the median pairwise distance as the bandwidth, as well as bandwidths scaled by factors from $0.001 \times$ up to $1000 \times$ relative to the median. The scores were normalized such that the maximum MMD score for each bandwidth result is 1, allowing us to observe trends in the scores independently of their absolute values.

The results are shown in Figure \ref{fig:bandwidth_score_norm}. A reliable metric is expected to show a monotonic increase in score as the severity of the degradation increases, because higher KAD scores should correlate with worse audio quality. When the median pairwise distance is used as the bandwidth, the MMD scores consistently respect the expected monotonic trend across all degradations tested. This suggests that the median heuristic provides stable and meaningful results. Bandwidths scaled by $10 \times$ and $100 \times$ the median also produced reliable results, maintaining the monotonic relationship. However, smaller bandwidths ($0.001 \times$ to $0.1 \times$ resulted in scores that drop too quickly, diminishing the discriminative power of the metric. Larger bandwidths ($1000 \times$ the median) tend to flatten the scores, reducing sensitivity to dissimilarities.

These initial experiments suggest that the median bandwidth heuristic is a robust and effective choice for calculating MMD scores. While certain scaled bandwidths may also be viable, the median heuristic avoids nonsensical results, ensuring stable and interpretable evaluations without additional tuning. Therefore, we recommend its use for kernel bandwidth selection in this context.

\section{Bias of FAD}
\label{sec:appendix_fad_bias}

In order to analyze the bias of FAD, we revisit Equation \ref{eq:FAD}:

\begin{center}
    $\text{FAD}^2(X,Y) = \|\mu_X - \mu_Y\|_2^2 \;+\; \text{tr}\left(\Sigma_X + \Sigma_Y - 2\sqrt{\Sigma_X \Sigma_Y}\right).$
\end{center}

For finite samples, $\mu_X, \mu_Y$ and $\Sigma_X, \Sigma_Y$ are replaced by their sample estimates $\hat{\mu}_X, \hat{\mu}_Y$ and $\hat{\Sigma}_X, \hat{\Sigma}_Y$. This introduces bias due to finite sample effects.

The first term, $\|\mu_X - \mu_Y\|^2$, is unbiased since the sample means $\hat{\mu}_X$ and $\hat{\mu}_Y$ are unbiased estimators of the true means $\mu_X$ and $\mu_Y$. Thus,
\begin{center}
    $\mathbb{E}[\|\hat{\mu}_X - \hat{\mu}_Y\|^2] = \|\mu_X - \mu_Y\|^2.$
\end{center}

The second term, $\text{tr}\left(\Sigma_X + \Sigma_Y - 2\sqrt{\Sigma_X \Sigma_Y}\right)$, is affected by finite sample sizes. The sample covariance matrix $\hat{\Sigma}$ is a biased estimator of the true covariance $\Sigma$:
\begin{center}
    $\mathbb{E}[\hat{\Sigma}] = \frac{N-1}{N} \cdot \Sigma$,
\end{center}

where $N$ is the sample size. Consequently, the expected value of the trace of the covariance matrices is:
\begin{center}
    $\mathbb{E}[\text{tr}(\hat{\Sigma}_X + \hat{\Sigma}_Y)] = \frac{N-1}{N} \cdot \text{tr}(\Sigma_X + \Sigma_Y).$
\end{center}

In the second term, $2 (\Sigma_X \Sigma_Y)^{1/2}$ is nonlinear in the covariances, and its exact bias is difficult to compute. However, to first-order approximation, the bias of FAD due to finite samples can be expressed as:
\begin{center}
    $\text{Bias}_{\text{FAD}} \approx \frac{1}{N} \cdot (\text{tr}(\Sigma_X) + \text{tr}(\Sigma_Y)) + \mathcal{O}\left(\frac{1}{N^2}\right).$
\end{center}

Here, the primary contribution to the bias arises from the underestimation of the covariance matrices, which scales inversely with the sample size \( N \).

\section{\texttt{kadtk}: KAD Toolkit Release}
\label{sec:kadtk_release}
We release a toolkit named \path{kadtk} that can calculate KAD scores from input audios. Given input audio directories for the ground-truth reference set and target evaluation set, the toolkit calculates the score and saves it to the given output file path in CSV format. The toolkit is written in Python and supports Pytorch and Tensorflow environments (refer to our Readme document for further details). It supports numerous models for embedding extraction: CLAP~\cite{clap_ms,clap_laion}, Encodec~\cite{encodec}, MERT~\cite{mert}, VGGish~\cite{vggish}, PANNs~\cite{panns}, OpenL3~\cite{openl3}, PaSST~\cite{passt}, DAC~\cite{dac}, CDPAM~\cite{cdpam}, Wav2vec2.0~\cite{wav2vec2.0}, HuBERT~\cite{hubert}, WavLM~\cite{wavlm}, and Whisper~\cite{whisper}. This covers a wide range of the audio domain including general environment sounds (sound effects, foley sounds, etc.), music, and speech. We also support FAD calculation for comparison.

The \path{kadtk} is released under the MIT License, allowing unrestricted use, modification, and distribution with proper attribution. The full license text is included in the repository. Some codes were brought from the FAD toolkit (fadtk) \cite{gui2024adapting,dcase2024,dcase2024report}. We sincerely thank the authors for sharing the code as an open source. Note that fadtk was also licensed under the MIT License. 

For the computational cost comparison in Section \ref{ssec:exp_compute}, we used \texttt{torch.linalg.eigval} and \texttt{torch.sqrt} to compute the covariance matrix of FAD, ensuring a fair comparison with KAD in a fully parallelized GPU setting. However, this approach is prone to accuracy issues, often leading to discrepancies with \texttt{fadtk}, which relies on \texttt{scipy.linalg.sqrtm} (available only for CPU computation). To ensure high FAD score accuracy at the cost of increased runtime, we used \texttt{scipy.linalg.sqrtm} for the perceptual alignment experiment (Section \ref{ssec:human_perception}) and in the released toolkit.




\end{document}